\documentclass[ aip, jcp, amsmath,amssymb,reprint]{revtex4-2}
\usepackage{amsmath}

\usepackage{graphicx}
\usepackage{dcolumn}
\usepackage{bm}

\usepackage[utf8]{inputenc}
\usepackage[T1]{fontenc}
\usepackage{mathptmx}
\usepackage{etoolbox}
\usepackage[colorlinks=true, allcolors=blue]{hyperref}

\usepackage[FIGBOTCAP]{subfigure}

\begin{document}

\title{Enhanced photoisomerization with hybrid metallodielectric cavities based on mode interference}

\author{Anael Ben-Asher}
\email{anael.benasher@uam.es}
\affiliation{Departamento de Física Teórica de la Materia Condensada and Condensed Matter Physics Center (IFIMAC), Universidad Autónoma de Madrid, E28049 Madrid, Spain}

\author{Thomas Schnappinger}%
\email{thomas.schnappinger@fysik.su.se}
\affiliation{Department of Physics, Stockholm University, Albanova University Centre, SE-106 91 Stockholm, Sweden}

\author{Markus Kowalewski}%
\email{markus.kowalewski@fysik.su.se}
\affiliation{Department of Physics, Stockholm University, Albanova University Centre, SE-106 91 Stockholm, Sweden}

\author{Johannes Feist}
\email{johannes.feist@uam.es}
\affiliation{Departamento de Física Teórica de la Materia Condensada and Condensed Matter Physics Center (IFIMAC), Universidad Autónoma de Madrid, E28049 Madrid, Spain}

\begin{abstract}
The ability to control chemical reactions by coupling organic molecules to confined light in a cavity has recently attracted much attention. While most previous studies have focused on single-mode photonic or plasmonic cavities, here we investigate the effect of hybrid metallodielectric cavities on photoisomerization reactions. Hybrid cavities, which support both photonic and plasmonic modes, offer unique opportunities that arise from the interplay between these two distinct types of modes. Specifically, we demonstrate that interference in the spectral density due to a narrow photonic mode and a broad plasmonic mode that are coupled to each other enables hybrid cavities to provide an energy-selective Purcell effect. This effect enhances electronic relaxation only to the desired molecular geometry, providing the ability to increase the yield of photoisomerization reactions. As a test case, we study the asymmetric proton transfer reaction in the electronic excited state of 3-aminoacrolein. Our results, which are robust for a range of realistic cavity parameters, highlight the advantages of hybrid cavities in cavity-induced photochemical processes.
\end{abstract}

\maketitle

\section{Introduction}
The coupling of organic molecules to confined light in a cavity and the formation of hybrid light-matter states, known as polaritons, enable the manipulation of both light and matter. In the past decade, their potential to tailor chemical reactions~\cite{feist2018polaritonic,hertzog2019strong,herrera2020molecular,Ebbesen2023-fd,Bhuyan2023-se,Kowalewski16jcp}, energy transport~\cite{balasubrahmaniyam2023enhanced,sokolovskii2023multi,sandik2024cavity}, lasing~\cite{kasprzak2006bose,kena2010room,Arnardottir2020-pn}, and photon non-linearities~\cite{toninelli2021single} has been extensively studied. However, conventional photonic cavities, such as Fabry–P\'erot resonators~\cite{gerard1996quantum}, primarily enable collective light-matter coupling involving many molecules~\cite{lidzey1999room}. This delocalized coupling limits their ability to influence properties at the level of individual molecules~\cite{galego2015cavity,cwik2016excitonic}. In contrast, plasmonic nanocavities overcome this limitation by providing extreme sub-wavelength confinement, enabling significant single-molecule coupling strengths, and offering a promising platform for controlling individual molecular properties~\cite{torma2014strong,chikkaraddy2016single,zhang2017sub,kongsuwan2018suppressed,ojambati2019quantum,baumberg2019extreme,heintz2021fewmolecule,li2022room}.

Recently, several studies~\cite{felicetti2020photoprotecting,wellnitz2020collective,torres2021molecular,wellnitz2021quantum} have shown that the coupling of a molecular electronic excitation to a plasmonic mode, which features high loss due to metal absorption, can tailor molecular photorelaxation and thus affect the molecular structure and dynamics. This phenomenon is attributed to the Purcell effect~\cite{Purcell,drexhage1970influence,kuhn2006enhancement,anger2006enhancement,kinkhabwala2009large}, wherein the cavity accelerates the molecular spontaneous emission rate by facilitating higher coupling to the free-space electromagnetic environment, as determined by the loss rate of the cavity. 
Thus, the Purcell effect, leveraging the high loss of the plasmonic mode, does not require the system to be in the strong coupling regime, where the light-matter coupling strength matches or exceeds the individual relaxation rates of the cavity and molecular excitation, but occurs in the weak coupling regime. This makes Purcell-induced reactions more feasible for experimental realization.

In contrast to many previous studies that have focused on one-mode plasmonic cavities~\cite{felicetti2020photoprotecting,wellnitz2020collective,torres2021molecular,wellnitz2021quantum}, here we investigate the effect of a hybrid metallodielectric cavity, which involves both plasmonic and photonic modes. These (at least) two-mode cavities have gained significant theoretical~\cite{yang2011hybrid,peng2017enhancing,gurlek2018manipulation,thakkar2017sculpting,ben2023non} and experimental~\cite{barth2010nanoassembled,luo2015chip,cui2015hybrid,son2024strong} interest in recent years, as they combine the low-loss properties of photonic microcavities with the highly localized fields of plasmonic modes, enabling novel functionalities. In this work, we explore their impact on the Purcell-induced photoisomerization reaction. 
In general, photoisomerization~\cite{sarto2022fundamentals} is a process in which a molecule absorbs light, promoting it to an excited electronic state, followed by isomerization to a different geometric configuration, and a subsequent relaxation back to the ground state. 
Such a relaxation can either happen via conical intersection on the sub-picosecond
timescale~\cite{Schnappinger22cc} or via spontaneous decay on a nanosecond timescale~\cite{sarto2022fundamentals}.
However, the spontaneous emission rate can be significantly accelerated via the Purcell effect by coupling the electronic transition to the cavity modes. Moreover, by selectively enhancing the decay rate to favor one geometry over another, the yield of the photoisomerization reaction can be increased. This selectivity is achieved through the energy selectivity of the cavity.

We propose that hybrid cavities provide a more energy-selective Purcell effect than single-mode cavities, enhancing the geometric selectivity of relaxation from the excited state and increasing the yield of photoisomerization reactions. As an illustrative case study, we investigate 3-aminoacrolein~\cite{Le-De2024-ff}, a model system for asymmetric proton transfer reactions in an excited electronic state. This model reaction, which demands high energy selectivity, highlights the advantages of hybrid cavities in controlling photoisomerization at the single-molecule level.
The remainder of this paper is organized as follows. In Section 2, we present the theoretical foundation of our work, explaining why hybrid cavities offer enhanced energy selectivity and detailing how the photoisomerization reaction can be theoretically studied, including an expression for the rate constant of cavity-mediated population transfer. Section 3 introduces the molecular case study, describing the computational methodology and the selected cavity parameters. In Section 4, we present the results that demonstrate the impact of the hybrid cavity on photoisomerization dynamics. Finally, we summarize our findings.

\section{Theory}
\subsection{Multimode Purcell enhancement}

The Purcell enhancement is determined in the weak coupling regime by the spectral density of the electromagnetic modes of the cavity, $J(\omega)$, at the transition frequency $\omega$, where a higher spectral density corresponds to a stronger interaction with the cavity and thus a faster relaxation~\cite{breuer2002theory}.
The energy gap between the electronic excited state and the electronic ground state dictates the transition frequency of the molecular excitation, given by: $V_e(X) - V_g(X)$, where $V_g(X)$ and $V_e(X)$ are the potential energy surfaces of the ground and excited state as a function of the reaction coordinate $X$. Thus, when $J(\omega)$ varies strongly with $\omega$, it provides a high selectivity in the spectral domain, which allows the possibility of selective coupling to certain regions on the potential energy surfaces, e.g., a specific molecular geometry. As a result, Purcell enhancement can favor specific relaxation pathways, thus influencing isomerization dynamics. 

Achieving both strong Purcell enhancement and energy selectivity in single-mode cavities can be challenging. The spectral density of a single cavity mode can be characterized by a Lorentzian function~\cite{grynberg2010introduction,medina2021few}:
\begin{equation}\label{1mod}
\text{J}_{\text{1 mode}}(\omega) = \frac{g^2 \Gamma}{2\pi[(\omega_c - \omega)^2 + \Gamma^2 / 4]},
\end{equation}
where the linewidth $\Gamma$ represents the decay rate of the cavity mode. Here, $\omega_c$ is the frequency of the cavity mode, and $g$ is its coupling strength to the molecule.  Typically, a larger $g$ is associated with a larger $\Gamma$~\cite{chikkaraddy2016single,zhang2017sub,kongsuwan2018suppressed,ojambati2019quantum}, implying that a higher Purcell enhancement comes at the cost of broader spectral density, which reduces selectivity.
Furthermore, the molecular enhanced decay is limited to half the cavity's loss rate, $\Gamma/2$, emphasizing the importance of a high $\Gamma$ for achieving strong Purcell enhancement~\cite{bozhevolnyi2016fundamental}.

\begin{figure}[t]
   \includegraphics[width=0.9\columnwidth, angle=0,scale=1,
draft=false,clip=true,keepaspectratio=true]{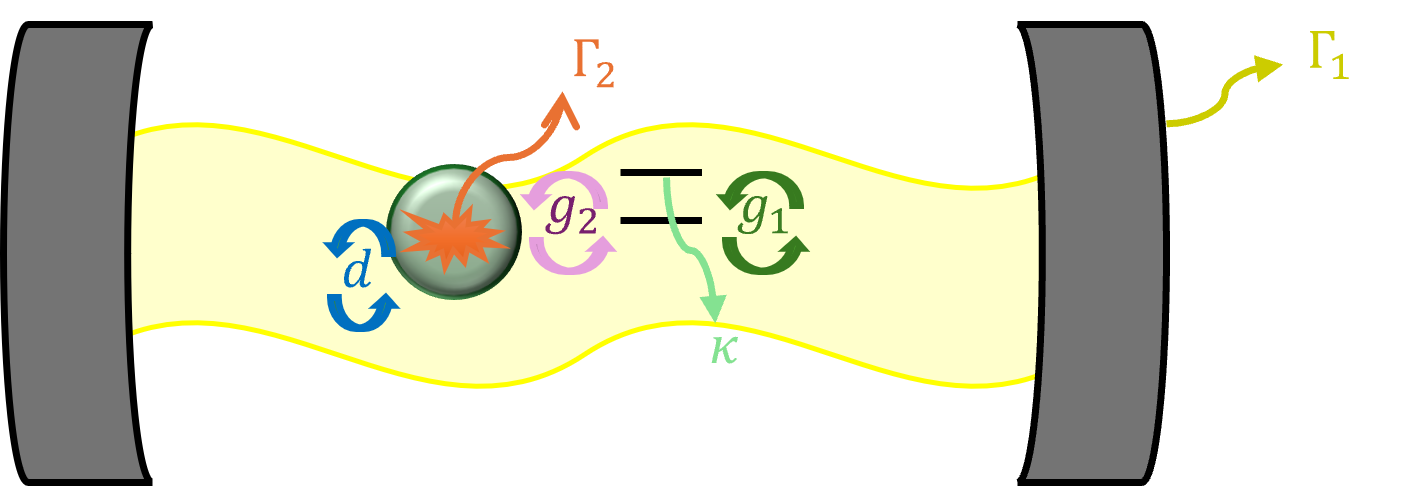}
\caption[]{Scheme of the hybrid metallodielectric cavity setup combining a Fabri P\'{e}rot photonic mode with a plasmonic mode and interacting with a matter excitation described by a two-level system.}
\label{fig:1}
\end{figure}

Consequently, we propose using a hybrid metallodielectric cavity setup, sketched in \autoref{fig:1}, which combines two electromagnetic modes: one plasmonic and one photonic, to achieve both high Purcell enhancement and energy selectivity. The plasmonic mode provides significant coupling to the molecule, while selectivity is achieved through interference between the two modes.
The spectral density of this hybrid cavity can be described as~\cite{medina2021few}:
\begin{equation}
\text{J}_{\text{2 mode}}(\omega) = \frac{1}{\pi} \mathrm{Im} \left\{ \vec{g}^T \frac{1}{\mathbf{H}_{\text{2 mode}} - \omega} \vec{g} \right\},
\end{equation}
where $\vec{g} = \{ g_1, g_2 \}$ and
\begin{equation}
\mathbf{H}_{\text{2 mode}} = \begin{pmatrix} \omega_1 - \frac{i}{2} \Gamma_1 & d \\ d & \omega_2 - \frac{i}{2} \Gamma_2 \end{pmatrix}.
\end{equation}
Here, $\omega_1$ and $\omega_2$ are the frequencies of the photonic and plasmonic modes, $\Gamma_1$ and $\Gamma_2$ are their decay rates, $g_1$ and $g_2$ represent their coupling strengths to the molecule, and $d$ is the coupling between the two modes originating from the interaction between the electric field of the photonic mode and the plasmonic dipole moment. For $d = 0$, the spectral density reduces to the sum of two Lorentzians. However, for non-zero $d$, asymmetric lineshapes emerge.
In the regime relevant to hybrid cavities, where the coupling strength and decay rate of the photonic mode are significantly smaller than those of the plasmonic mode, that is, $\Gamma_2\gg\Gamma_1$ and $g_2\gg g_1$, the spectral density features two asymmetric peaks with the same amplitudes but different widths and a dip between them. Specifically, when $g_1 = 0$ and $\Gamma_1 = 0$, the spectral density simplifies to:
\begin{equation}\label{2mod}
\text{J}_{\text{2 mode}}(\omega) = \frac{g_2^2 \Gamma_2}{2\pi \left[ (\omega_2 - \frac{d^2}{\omega_1 - \omega} - \omega)^2 + \frac{\Gamma_2^2}{4} \right]}
\end{equation}
for $\omega \neq \omega_1$, featuring two maxima when $\omega_2 - \frac{d^2}{\omega_1 - \omega} - \omega=0$, and is zero at $\omega = \omega_1$. By selecting appropriate values for $\omega_1$, $\omega_2$, $\Gamma_2$, $g_2$, and $d$, we engineer one of the spectral density peaks to be very narrow, allowing for highly energy-selective Purcell enhancement at the right frequency.

\subsection{The rate constant for cavity-mediated population transfer}
We study the Purcell-induced photoisomerization reaction by solving a set of rate equations that describe the population transfer between different vibrational states, representing different molecular isomers, in the molecular electronic ground state. 
These states are given by the eigenfunctions $|\Phi_n\rangle$  of $\hat{H}_g=\hat{T}+V_g(X)$, the nuclear Hamiltonian in the electronic ground state within the Born–Oppenheimer approximation, where $\hat{T}$ is the kinetic energy operator along the reaction coordinate $X$. 
The transfer of population from $|\Phi_i\rangle$ to $|\Phi_f\rangle$, mediated by excitation to the electronic excited state and decay through cavity losses, is a second-order process. Therefore, its rate constant $k_{i\to f}^{ESPT}$ is given by the Kramers-Heisenberg formula~\cite{loudon2000quantum}. We use its non-Hermitian (NH) form~\cite{moiseyev2004non,ben2017boomerang,ben2020quantum} in order to account for cavity losses and the spontaneous decay rate of the molecule, $\kappa$. 

The hybrid cavity coupled to a molecular electronic excitation corresponds to the following NH Hamiltonian:
 \begin{multline}
  \label{H}
   \hat{H}_0 = \hat{H}_g+\bigg(\hat{H}_e-\hat{H}_g-i\frac{\kappa}{2}\bigg)\sigma_+\sigma_-+\sum_{n=1,2}\big[\tilde{\omega}_n a^\dagger_n a_n \\
   + g_n D(X)(\sigma_+a_n+a^\dagger_n\sigma_-)\big]+d(a^\dagger_1a_2+a^\dagger_2a_1),
 \end{multline}
where the rotating-wave approximation is applied. This Hamiltonian is the effective NH Hamiltonian~\cite{roccati2022non} arising from the Lindblad master equation~\cite{breuer2002theory,lindblad1976generators}, which is commonly used to describe lossy systems.
Here, $\sigma_-$ ($\sigma_+$) and $a_n$ ($a_n^\dagger$) denote the annihilation (creation) operators of the molecular exciton and the two optical modes, respectively; $\hat{H}_e=\hat{T}+V_e(X)$ is the nuclear Hamiltonian in the excited electronic state;  $\tilde{\omega}_n=\omega_n-i\frac{\Gamma_n}{2}$ are the complex energies of the cavity modes; and $D(X)$ is the geometry-dependent transition dipole moment between the ground and the excited electronic states.  The initial and final states of the population transfer correspond to the eigenstates of the zeroth optical excitation of $\hat{H}_0$. In these states, the cavity modes are in their corresponding vacuum state $|0\rangle$. They are given by $|i\rangle=|\Phi_i\rangle\otimes|0\rangle=|\Phi_i,0\rangle$ and $|f\rangle=|\Phi_f\rangle\otimes|0\rangle=|\Phi_f,0\rangle$ with the eigenenergies $E_i$ and $E_f$. 
However, the states that mediate population transfer correspond to the first optical excitation manifold of $\hat{H}_0$ and are hybrid light-matter states, which we here will denote generically as polaritons without distinguishing between weak and strong coupling regimes. While the zeroth optical excitation of $\hat{H}_0$ is described by a Hermitian Hamiltonian, its first optically excited manifold is given by an NH Hamiltonian due to the difference in decay rate between the cavity modes ($\Gamma_1,\Gamma_2$) and the emitter ($\kappa$). 
Therefore, to emphasize their different inner product~\cite{moiseyev2011non}, the notation $|r)$ and $(r|$ is used, rather than $|r\rangle$ and $\langle r|$, for the right and left eigenstates of the first optical excitation manifold of $\hat{H}_0$. 
Since $\hat{H}_0$ is complex symmetric, $|r)$ is the same as $|r\rangle$, while $(r|$ is the complex conjugate of $\langle r|$. 
Moreover, these eigenstates are associated with complex eigenvalues $\tilde{E}_r$~\cite{moiseyev2011non}. 

The rate constant $k_{i\to f}^{ESPT}$ for the population transfer mediated by laser excitation and subsequent decay is then given by: 
 \begin{eqnarray}\label{kes}
k_{i\to f}^{ESPT}=\sum_k\bigg|\sum_r \frac{ \langle0,\Phi_f|\hat{V}_d^{(k)}|r) (r|\Omega D(X)\sigma_+|\Phi_i,0\rangle}{\tilde{E}_r-E_i-\hbar\omega_L}\bigg|^2,
\end{eqnarray}
where the index $k$ sums over all possible decay channels.
The matrix element $(r|\Omega D(X)\sigma_+|\Phi_i,0\rangle$ describes the laser excitation of the initial state into the first optical excitation manifold of $\hat{H}_0$ through the molecular transition dipole $D(x)$ where $\Omega$ is the laser's field strength, and the denominator of \autoref{kes}, which does not diverge due to the complex value of $\tilde{E}_r$, describes the energy match with the laser frequency, $\omega_L$.
In addition, the matrix element $\langle0,\Phi_f|\hat{V}_d^{(k)}|r)$ describes the subsequent relaxation from the optically excited manifold to the final state through the channel $k$. 
Following Refs.~\citenum{moiseyev2004non,ben2017boomerang,ben2020quantum}, we use $\sqrt{\kappa}\sigma_-$ for the spontaneous emission of the molecule, and $\sqrt{\Gamma_1}a_1$ and $\sqrt{\Gamma_2}a_2$ for the two cavity losses as decay operators $\hat{V}_d^{(k)}$. 
The derivation of \autoref{kes} with these decay operators from the Lindblad master equation using perturbation theory is presented in the appendix. 

In \autoref{kes}, two roles for the cavity setup are taken into account. First, the decay rate from the optically excited manifold to the electronic ground state is enhanced by opening additional decay channels beyond molecular spontaneous emission. 
Second, the eigenstates in the optically excited manifold of $\hat{H}_0$ are modified due to the formation of polaritons. If these polaritonic effects are neglected, the rate constant simplifies to:
\begin{eqnarray}\label{kesNP}
k_{i\to f}^{ESPT (nonP)}=\bigg|\sum_e \frac{\sqrt{\gamma_{ef}}\langle\Phi_f|D(X)|\phi_e\rangle\langle\phi_e|\Omega D(X)|\Phi_i\rangle}{E_e-i\kappa/2-E_i-\hbar\omega_L}\bigg|^2.
\end{eqnarray}
Here, we assume that only the eigenstates of $\hat{H}_e$, $|\phi_e\rangle$, with eigenvalues $E_e$, and not the polaritons $|r)$, which are the eigenstates of the first optical excitation manifold of $\hat{H}_0$, mediate the population transfer.  In this case, the effect of the cavity setup on the decay to the electronic ground state is implicit and treated perturbatively~\cite{breuer2002theory}, assuming that the cavity setup can be described as a Markovian bath weakly coupled to the molecular excitation. As a result, the cavity-enhanced decay from $|\phi_e\rangle$ to $|\Phi_f\rangle$  is given by the matrix element $\langle\Phi_f|D(X)|\phi_e\rangle$ of the molecular transition dipole through which the cavity is coupled, multiplied by the square root of the rate $\gamma_{ef}=2\pi J(E_e-E_f)$ where $J(\omega)$ is the spectral density of the cavity setup.
When valid, \autoref{kesNP} allows the calculation of cavity-mediated population transfer without explicitly considering the cavity degrees of freedom, simplifying the treatment of multiple modes. The laser excitation in \autoref{kesNP} is treated similarly to \autoref{kes}, using the molecular transition dipole. The molecular spontaneous decay rate $\kappa$ is used to broaden the electronic excited states, preventing divergence of \autoref{kesNP}.

\subsection{The molecular steady state}
Our description of the dynamics of the cavity molecular system is based on coupled rate equations. To determine the rate constants $k_{i\to f}^{ESPT}$, we assign the rate equation to each vibrational state in the molecular electronic ground state:
\begin{eqnarray}
\frac{dP_n(t)}{dt}=\sum_{m\neq n} k^{tot}_{m\to n}P_m(t)-P_n(t)\sum_{m\neq n} k^{tot}_{n\to m}
\end{eqnarray}
where $P_n(t)$ is the population in the state $\Phi_n$, and $k^{tot}_{m\to n}$ is the total rate constant for population transfer from the state $\Phi_m$ to state $\Phi_n$, summing the rate constant $k_{m\to n}^{ESPT}$ derived above with the vibrational relaxation rate constant $k_{m\to n}^{VR}$. The last  accounts for the effect of the nuclear degrees of freedom neglected when computing $k_{m\to n}^{ESPT}$ considering the reduced reaction coordinate, and is given by:
\begin{eqnarray}
k_{m\to n}^{VR}=\gamma_0|\langle\Phi_n|X|\Phi_m\rangle|^2
\end{eqnarray}
for $E_m>E_n$, while it is zero for $E_m<E_n$. The matrix element $\langle\Phi_n|X|\Phi_m\rangle$ describes the transition dipole between the two states and $\gamma_0=\frac{1}{ps}$. 
We find the photostationary state of this setup by locating the populations ${P_n(t\to\infty)}$ for which $\frac{dP_n(t\to\infty)}{dt}=0$ for some $n$ and which obey $\sum_n P_n(t\to\infty)=1$. By determining which state is associated with each isomer of the molecule, the photoisomerization process can be studied.

\section{The molecular case study}

\begin{figure}[t]
   \includegraphics[width=1\columnwidth, angle=0,scale=1,
draft=false,clip=true,keepaspectratio=true]{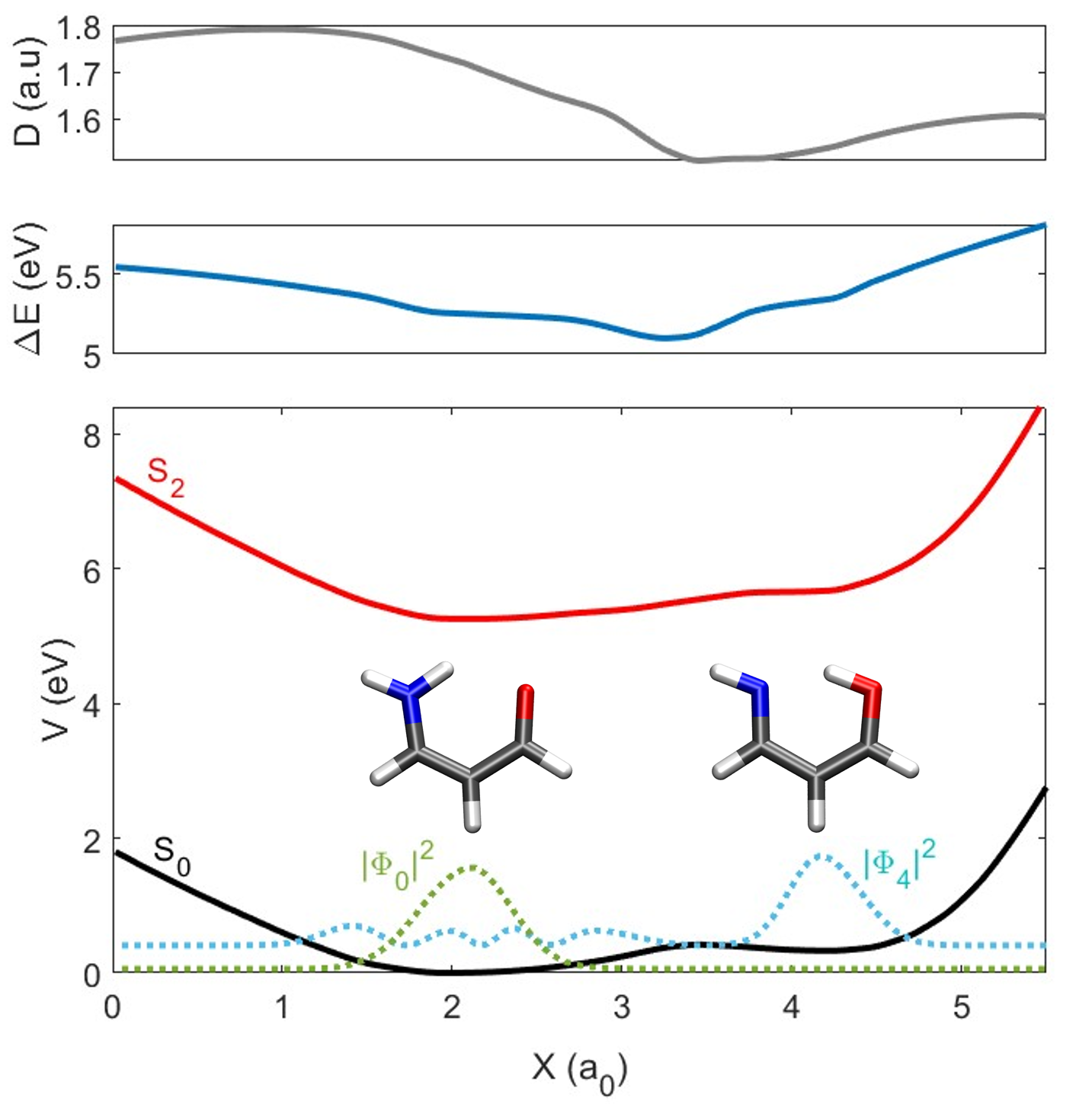}
\caption[]{Lower panel - the potential energy curves of the electronic ground (S$_0$) and second excited (S$_2$) states of (Z)-3-aminoacrylaldehyde as a function of the proton transfer reaction coordinate $X$ defined by the IRC (see main text). The vibrational wavefunctions associated with the ground state of each isomer are also plotted. Middle and upper panels- the excitation energy and the transition dipole between these two electronic two states as a function of $X$.}
\label{fig:2}
\end{figure}

As a case study, we consider the asymmetric proton transfer reaction between the nitrogen atom and the oxygen atom in the second electronic excited state S$_2$ of (Z)-3-aminoacrylaldehyde~\cite{Le-De2024-ff}.
The chosen molecular model represents an asymmetric variant of the well-studied symmetric proton transfer in (Z)-malondialdehyd~\cite{Isaacson1975-ew,Karlstrom1975-kr,Ruf1988-ul}.
The reaction of (Z)-3-aminoacrylaldehyde can be well approximated in a lower-dimensional subspace~\cite{Ruf1988-ul,Zauleck2016-ab,Zauleck2018-rp}.
Both minimum structures (depicted in \autoref{fig:2}), the transition state of the proton transfer reaction, and the corresponding intrinsic reaction coordinate (IRC) were calculated in the electronic ground state S$_0$ at the $\omega$B97XD/aug-CC-PVDZ~\cite{Chai2008-wv,Kendall1992-wu} level of theory using Gaussian 16 Rev.C.01 program package~\cite{g16}. 
The potential energy curve for the optically accessible S$_2$ state and the corresponding transition dipole moment are obtained using linear-response time-dependent density functional theory within the Tamm-Dancoff approximation~\cite{Hirata1999-so} with 10 roots. 
All electronic structure calculations were performed in a reproducible environment using the Nix package manager together with NixOS-QChem \cite{nix} (commit f803c222) and Nixpkgs (nixpkgs, 23.05, commit 5550a85a).
The last point of the IRC in both directions and the corresponding minima were used to extrapolate the potential energy curve to obtain a bound potential for S$_2$ and S$_0$. These potentials are plotted in the lower panel of \autoref{fig:2} as a function of the reaction coordinate $X$ defined by the IRC, which describes the proton transfer reaction between nitrogen and oxygen atoms. 
The kinetic energy operator along the reaction coordinate $X$ is constructed within the G-Matrix formalism~\cite{Podolsky:1928,Wilson:1980,Schaad:1989} as described in Ref.~\citenum{Thallmair:2016}.
It reads
\begin{equation}
 \hat{T} \approx - \frac{\hbar^{2}}{2} \frac{\partial}{\partial X}  G(X) \frac{\partial}{\partial X} 
\end{equation}
with the G-Matrix element $G(X)$ being computed via finite differences by displacement of the Cartesian molecular geometry along the internal coordinate.  
The vibrational states for the electronic ground state (S$_0$) and the excited state (S$_2$) along $X$, corresponding to the eigenstates of $\hat{H}_g$ and $\hat{H}_e$ (in \autoref{H}), were determined using the discrete variable representation (DVR) method and the one-dimensional particle-in-a-box basis functions. The chosen parameters, for which we obtain converged results, are as follows: box length $L= 6$~a.u.; and the number of basis functions $n=100$. 
The vibrational ground state in S$_0$, $\Phi_{0}(X)$,  presented in the lower panel of \autoref{fig:2}, is localized in the left well. 
After the molecule has been excited to the second excited electronic state, the excited-state wave packet can propagate between the left and right wells of the S$_2$ potential curve. The subsequent decay back to the ground electronic state S$_0$ can give rise to the product of the proton transfer reaction as the right well in the S$_0$ potential curve is populated.
Specifically, we are interested in transferring the population to the fourth vibrational state in S$_0$, $\Phi_{4}(X)$, which is localized mostly in the right well, as is also shown in the lower panel of \autoref{fig:2}. 

The enhancement of population transfer to $\Phi_{4}(X)$ by coupling the molecular excitation S$_0\to$S$_2$ to a cavity setup requires high energy selectivity, since both the excitation energy, given by the difference between the two potentials $\Delta E(X)=V_{\text{S}_2}(X)-V_{\text{S}_0}(X)$ (middle panel of \autoref{fig:2}), and the electronic transition dipole $D(X)$ (upper panel of \autoref{fig:2}), are similar between the two nuclear configurations. 
Thus, it fits the purpose of demonstrating the benefits of the hybrid metallodielectric cavity in enhancing photoisomerization reactions. 
The laser frequency used matches the highest Franck-Condon factor corresponding to $\omega_L=E_{0,e}-E_{0,g}=5.254$~eV, where $E_{0,e/g}$ are the lowest vibrational eigenvalues of the two electronic states. 
In addition, we consider $\kappa=10^{-7}\,\text{a.u.} =4\, \text{ns}^{-1}$ as the rate of spontaneous decay of the molecule. 
We set the coupling strength and the decay rate of the photonic mode to $g_1=0$ and $\Gamma_1=0$, which give the strongest destructive interference between the modes and, therefore, the highest energy selectivity. The dependence of the results on $g_1$ and $\Gamma_1$ is presented below.
The other parameters of the hybrid cavity, $\omega_1,\omega_2,\Gamma_2,g_2$ and $d$, are chosen using the constrained nonlinear optimization algorithm of MATLAB R2024a by looking for a maximum value of $P_{4}$ where $\Omega=10^{-5}$ a.u.$=5.142\frac{\text{V}}{\mu m}$. It is important to note that the field strength $\Omega$ includes the enhancement of the laser field by the cavity setup, which can be up to a few orders of magnitude in comparison to the original laser field~\cite{feist2020macroscopic}. 
The optimized values are summarized in \autoref{tab} and are feasible in accordance with the existing literature. 
Although we keep the plasmonic coupling strength $g_2$ within the weak coupling regime and up to $100$ meV as realized in Ref.~\citenum{chikkaraddy2016single,zhang2017sub,kongsuwan2018suppressed,ojambati2019quantum}, the coupling between the modes $d$ can be strong~\cite{garcia2024realization}. 
In addition, a high plasmonic decay rate equivalent to $~0.5$ eV was calculated for an aluminum sphere~\cite{torres2021molecular}. 
For comparison, we also optimize the results for a one-mode cavity, whose only parameters are $\omega_2,\gamma_2$ and $g_2$, and their optimized values are given in the second column of the \autoref{tab}.

\begin{table}[ht!]
{\caption{The optimization of the cavities' parameters to achieve the highest steady-state population in $\Phi_{4}(X)$ for the laser strength $\Omega=10^{-5}$. The optimal values for the two-mode cavity are given in the first column, and for the one-mode cavity, they are given in the second column.
}\label{tab}
}
\begin{center}
\begin{tabular}{c || c | c }
\hline
\hline
\multicolumn{1}{c||}{}&
\multicolumn{1}{c|}{two-mode optimal (eV)}
&
\multicolumn{1}{c}{one-mode optimal (eV)}\\
\hline

$\omega_1$ &     $5.351$ & -\\
\hline
$\omega_2$ &   $6.816$  & $5.276$ \\ \hline 
$\gamma_2$ &  $0.335$ & $0.448$\\ \hline 
$g_2$ & $0.046$ & $0.019$ \\  \hline
$d$ &  $0.164$ & -\\ \hline \hline
\end{tabular}
\end{center}
\end{table}

\section{Results and discussion}
\begin{figure}[t]
   \includegraphics[width=1\columnwidth, angle=0,scale=1,
draft=false,clip=true,keepaspectratio=true]{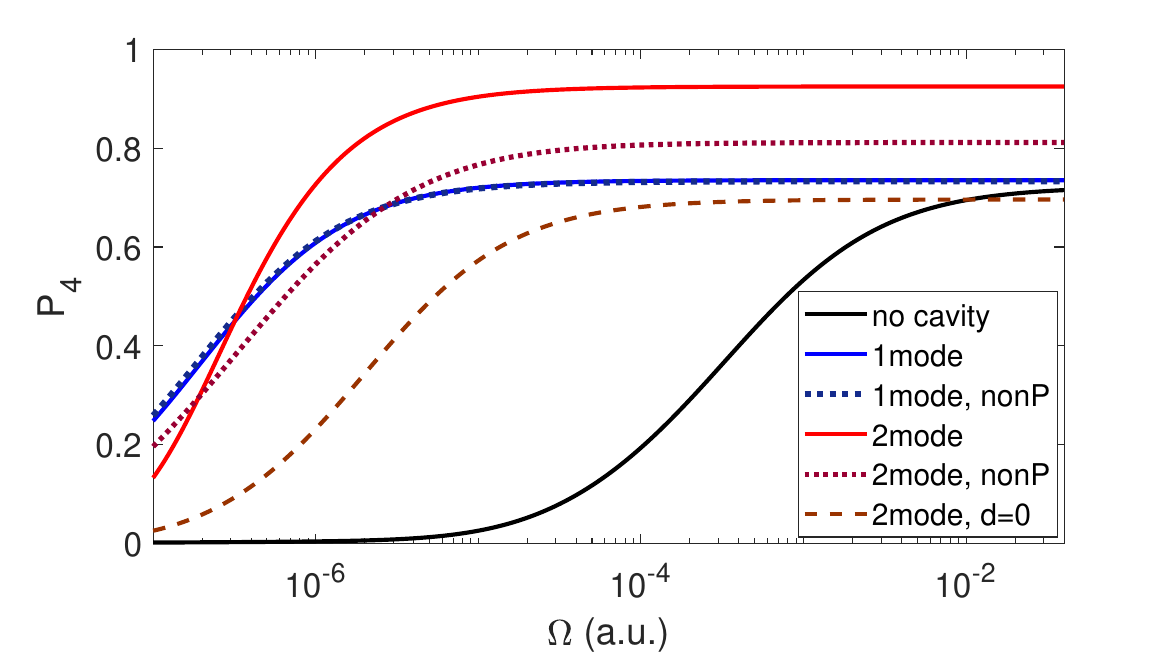}
\caption[]{The photo steady-state population of the vibrational state $\Phi_{4}$ in the electronic ground state (S$_0$), $P_{4}$, as a function of the laser field strength $\Omega$ for a setup of no cavity, 1-mode cavity, and 2-mode cavity. The parameters of the cavities were optimized to achieve a high population when $\Omega=10^{-5}$ a.u., as described in the main text.
The results obtained when neglecting the coupling between the two modes $d$ and when considering only the non-polaritonic effect of the two-mode cavity (\autoref{kesNP}) are also presented. }
\label{fig:3}
\end{figure}

The steady-state population of $\Phi_{4}(X)$, $P_4$, is presented in \autoref{fig:3} as a function of the laser strength $\Omega$ for both the optimized hybrid (two-mode) and one-mode cavity setups. For comparison, we also show the case without any cavity coupling, where spontaneous emission of the molecule solely governs the decay from the electronically excited state.
The dependence of $P_4$ on $\Omega$ reflects the interplay between the timescales of population transfer via the electronic excited state and the vibrational relaxation. The latter is fixed at $1$~picosecond in our calculations and remains independent of $\Omega$. For very small $\Omega$, vibrational relaxation dominates the dynamics, leading to a localization of the population in the vibronic ground state, with $P_4 \to 0$. As $\Omega$ increases, the transfer of population via the electronic excited state becomes more significant, facilitating the population of $\Phi_4(X)$ and increasing $P_4$. For sufficiently large $\Omega$, $P_4$ reaches a maximal constant value, indicating that the rate of population transfer through the electronic excited state exceeds the vibrational relaxation rate.
Comparing the results of the one-mode cavity and the two-mode cavity with the cavity-free case shows that the cavities accelerate population transfer to $\Phi_4(X)$, as they enable maximal population at lower $\Omega$ values than in the cavity-free setup. This acceleration arises from the Purcell effect, which enhances the spontaneous emission from S$_2$ to S$_0$. However, while the one-mode cavity achieves a maximal value of $P_4$ of approximately 70\%, similar to the cavity-free setup, the two-mode cavity increases this value to greater than 90\%. This enhancement highlights the selective effect of the hybrid cavity, as analyzed through the spectral densities below.

\begin{figure}[t]
   \includegraphics[width=1\columnwidth, angle=0,scale=1,
draft=false,clip=true,keepaspectratio=true]{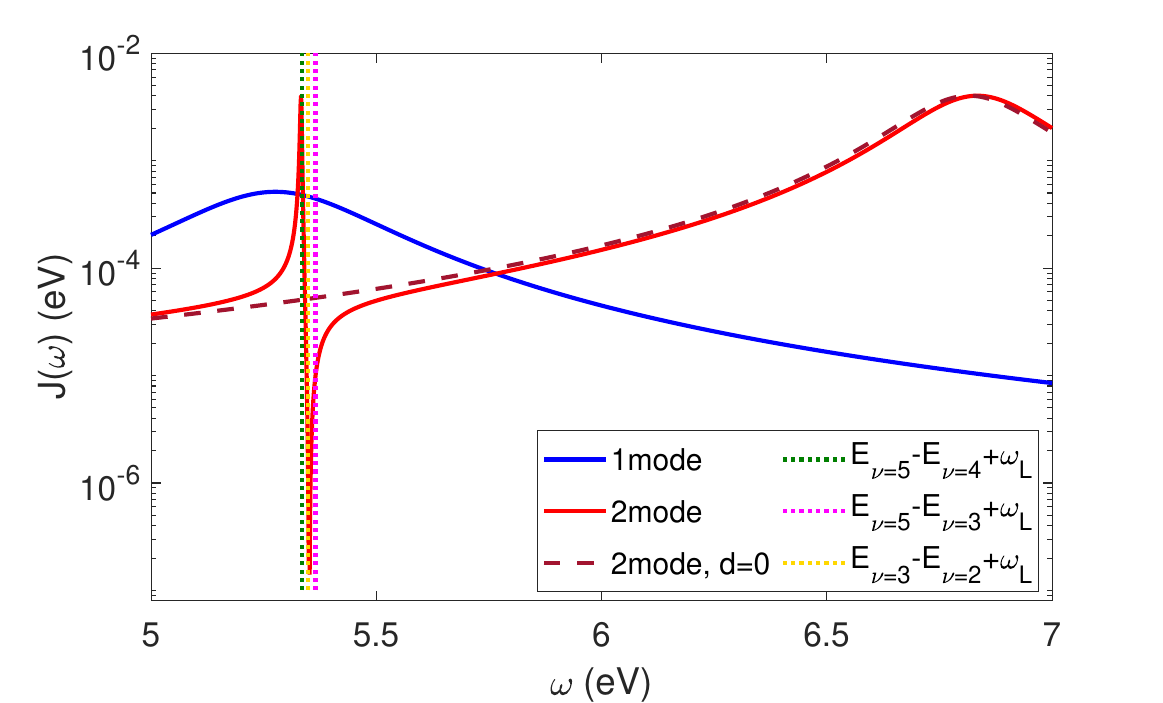}
\caption[]{The spectral densities of the one-mode and two-mode cavities corresponding to the results presented in \autoref{fig:3}. The vertical dashed lines indicate the energy gaps between different transitions and demonstrate the high selectivity achieved by the two-mode cavities. }
\label{fig:4}
\end{figure}

The spectral densities for the one-mode and two-mode cavity setups, J$_{1\text{mode}}(\omega)$ and J$_{2\text{mode}}(\omega)$, are shown in \autoref{fig:4} and are given by \autoref{1mod} and \autoref{2mod}, respectively. J$_{1\text{mode}}(\omega)$ exhibits a broad peak, whereas J$_{2\text{mode}}(\omega)$ features a narrower and asymmetric peak in the relevant frequency range for spontaneous emission from S$_2$ to S$_0$. This difference in peak linewidths indicates that the hybrid cavity offers a higher energy selectivity, which selectively enhances the decay from S$_2$ to $\Phi_4(X)$.
In the Markovian and weak-coupling regime, where the population in cavity modes is negligible, the effect of the cavities can be treated perturbatively.
Consequently, the Purcell-enhanced spontaneous decay rate depends on the value of the spectral density at the transition frequency, as taken into account in \autoref{kesNP}.
We present in \autoref{fig:3} in dotted lines the population $P_4$ obtained when replacing $k_{i\to f}^{ESPT}$ (\autoref{kes}) by $k_{i\to f}^{ESPT (nonP)}$ (\autoref{kesNP}) for the two cavity setups. 
For the one-mode cavity, whose coupling strength is smaller than that of the two-mode cavity, this approximate calculation reproduces the full calculation, while for the two-mode cavity, it does not. 
This discrepancy highlights the role of polariton formation and the exchange of energy between the molecule and cavity modes in the hybrid setup. Importantly, the higher $P_4$ predicted by the approximate calculation for the two-mode cavity reveals a strong connection between the line shape of the spectral density and the enhancement of the population, showing that the superior energy selectivity of the hybrid cavity setup is actually responsible for the increase in population transfer to $\Phi_4(X)$.

\begin{figure}[t]
   \includegraphics[width=1\columnwidth, angle=0,scale=1,
draft=false,clip=true,keepaspectratio=true]{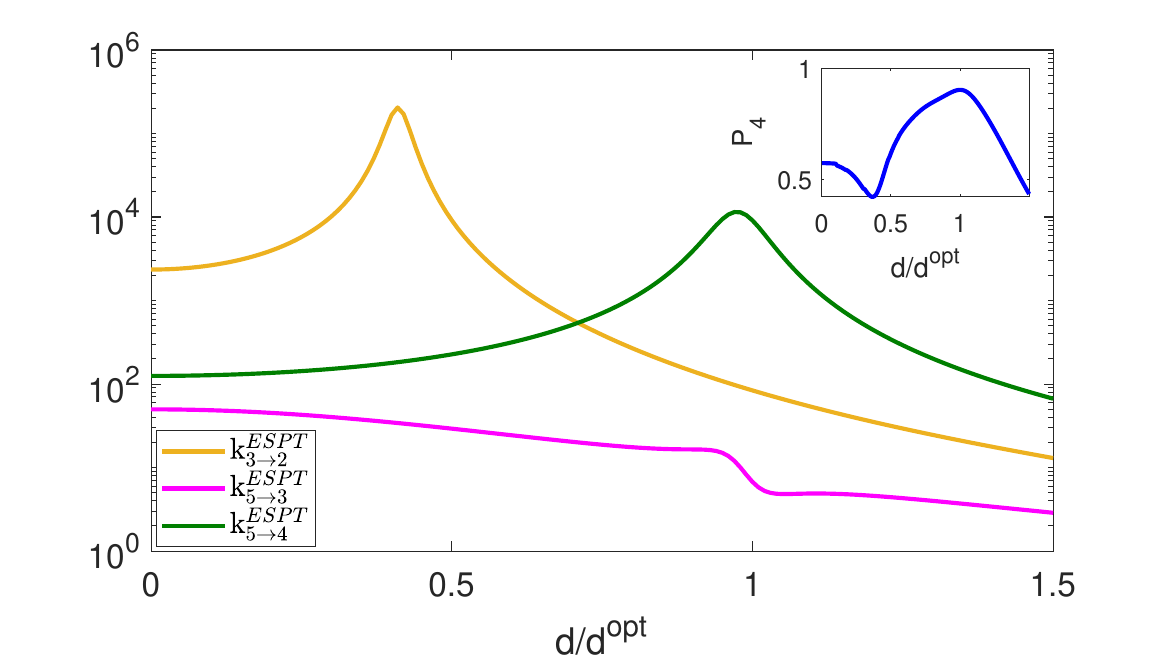}
\caption[]{The rate constants of competing transitions as a function of the coupling between the two modes $d$. The largest value for population $P_{4}$ is obtained for $d^{\mathrm{opt}}$ as shown in the inset.}
\label{fig:5}
\end{figure}

The narrow peak of J$_{2\text{mode}}$, which results in the high energy selectivity of the hybrid cavity, arises from the interference between the two modes due to their non-zero coupling, $d$, as discussed in the Theory section. In \autoref{fig:3}, the results for the two-mode cavity with $d=0$ are shown in a dashed line, where the corresponding spectral density is plotted in \autoref{fig:4} (also in a dashed line). Since $g_1=0$, this spectral density fits that of a one-mode cavity characterized by the parameters of the plasmonic mode, which is far detuned from the molecular excitation energies (\autoref{fig:4}). Consequently, this spectral density remains approximately constant in the relevant energy range for spontaneous emission from S$_2$ to S$_0$, leading to an unselective acceleration of emission and no increase in the asymptote of $P_4$ compared to the cavity-free case (\autoref{fig:3}).
To further investigate the role of $d$ and to highlight the importance of energy selectivity of the hybrid cavity setup in the enhancement of excited-state mediated population transfer, we compare the rate constants of the competing transitions, $k_{3\to 2}^{ESPT}, k_{5\to 3}^{ESPT}$ and $k_{5\to 4}^{ESPT}$, as functions of $d$ in \autoref{fig:5}. The vertical dotted lines in \autoref{fig:4} indicate the emission frequencies of these competing transitions, given by $E_i+\omega_L-E_f$ where $E_i$ and $E_f$ are the energies of the initial and final states, respectively, and $\omega_L$ is the laser frequency. Since these frequencies are very similar, the one-mode cavity and the setup for $d=0$ enhance these transitions unselectively, leading to the second highest populated state being $\Phi_{2}(X)$ due to the transitions $5\to3\to2$.
In contrast, the two-mode cavity selectively accelerates the transition $5\to4$ while suppressing the transitions $5\to3\to2$, thereby enhancing $P_4$ by carefully choosing $d,\omega_1$ and $\omega_2$ to tune the narrow peak's frequency (as determined from \autoref{2mod}). This high energy selectivity is demonstrated in \autoref{fig:5}, which shows that varying $d$, and thus varying the frequency corresponding to the maximum of the narrow peak in J$_{2\text{mode}}$, results in different transitions being accelerated. Note that higher values of $d$ correspond to lower frequencies for the narrow peak, which means that the maximal value of $k_{3\to 2}^{ESPT}$ whose transition energy exceeds that of $k_{5\to 4}^{ESPT}$, is obtained at smaller values of $d$.

\begin{figure}[t]
   \includegraphics[width=1\columnwidth, angle=0,scale=1,
draft=false,clip=true,keepaspectratio=true]{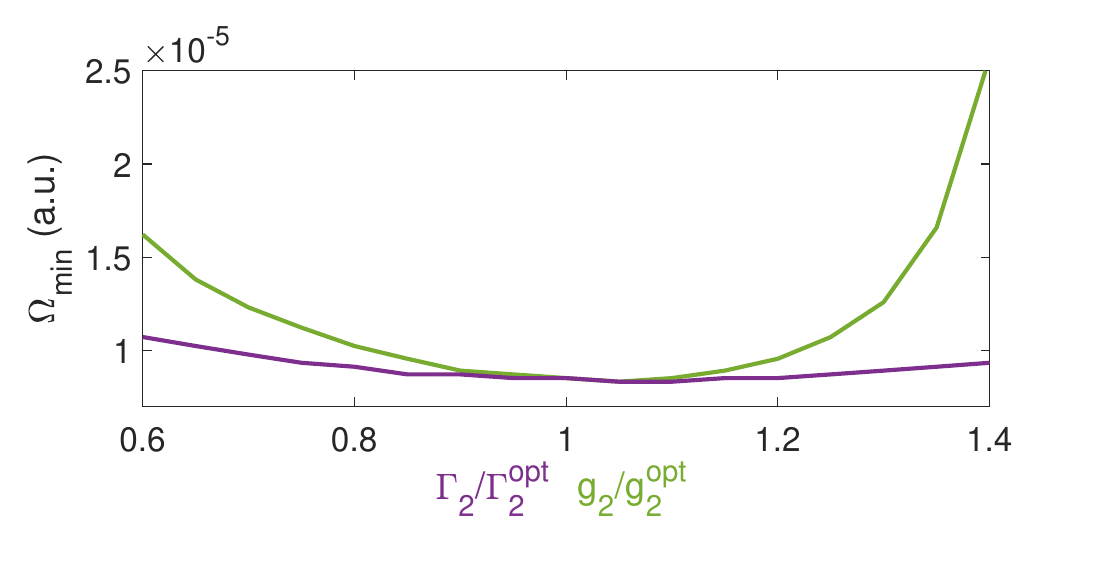}
\caption[]{The minimal laser field strength $\Omega_{min}$ for which $P_{4}>0.9$ as a function of the plasmonic decay rate $\Gamma_2$ and as a function of the plasmonic coupling strength  $g_2$. }
\label{fig:6}
\end{figure}

According to \autoref{2mod}, the parameters $d,\omega_L,\omega_1$ and $\omega_2$ determine the position of the maximum of J$_{2\text{mode}}(\omega)$ and are therefore responsible for achieving the required energy selectivity for population transfer to $\Phi_{4}(X)$. This is demonstrated in \autoref{fig:5} for variation in $d$.  In contrast, the parameters $g_2$ and $\Gamma_2$ dictate the amplitude of J$_{2\text{mode}}(\omega)$. Specifically, J$_{2\text{mode}}(\omega)$ scales linearly with $g_2$ for all $\omega$, while its scaling with $\Gamma_2$ is inversely proportional when $|\omega_2 - \frac{d^2}{\omega_1 - \omega} - \omega|\ll\Gamma_2/2$ and linear when $|\omega_2 - \frac{d^2}{\omega_1 - \omega} - \omega|\gg\Gamma_2/2$. 
In \autoref{fig:6} we present the minimal laser field strength $\Omega_{min}$ required for $P_{4}>0.9$ as a function of $g_2$ when $\Gamma_2=\Gamma_2^{opt}$ by the purple line and as a function of $\Gamma_2$ when $g_2=g_2^{opt}$ by the light green line. The parameters $g_2^{\mathrm{opt}}$ and $\Gamma_2^{\mathrm{opt}}$ were optimized to achieve the largest value for $P_{4}$ for the two-mode cavity setup and are given in \autoref{tab}.
Although the maximum value of $P_4$ remains relatively insensitive to variations in $g_2$ or in $\Gamma_2$, it varies with $\Omega_{min}$, i.e., the minimum laser power required to ensure that population transfer via the electronic excited state overcomes vibrational relaxation.
For a low amplitude of J$_{2\text{mode}}(\omega)$, the acceleration of spontaneous emission by the cavity is weaker, necessitating a stronger laser to achieve maximal $P_4$. However, if the amplitudes become too high, the spectral density description of the cavity-enhanced decay rate becomes invalid.
Thus, \autoref{fig:6} indicates that there exists an optimal range for $g_2$ and $\Gamma_2$  where population transfer via the excited state occurs most rapidly.

\begin{figure}[t]
\includegraphics[width=0.5\textwidth, angle=0,scale=1,
draft=false,clip=true,keepaspectratio=true]{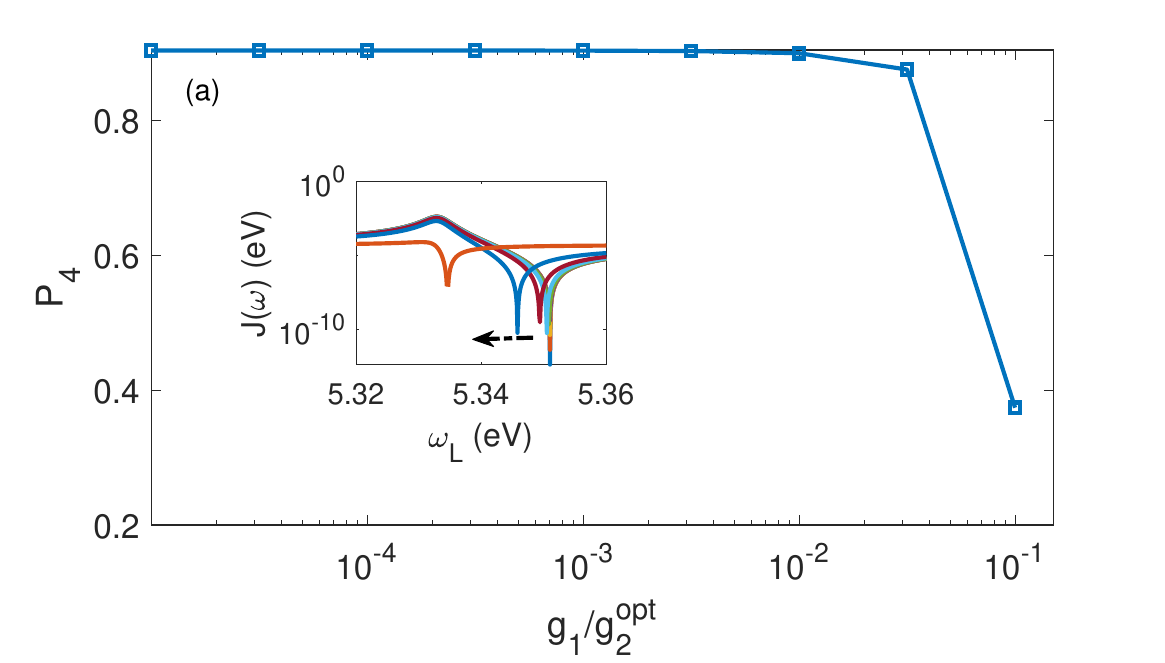}
\includegraphics[width=0.5\textwidth, angle=0,scale=1,
draft=false,clip=true,keepaspectratio=true]{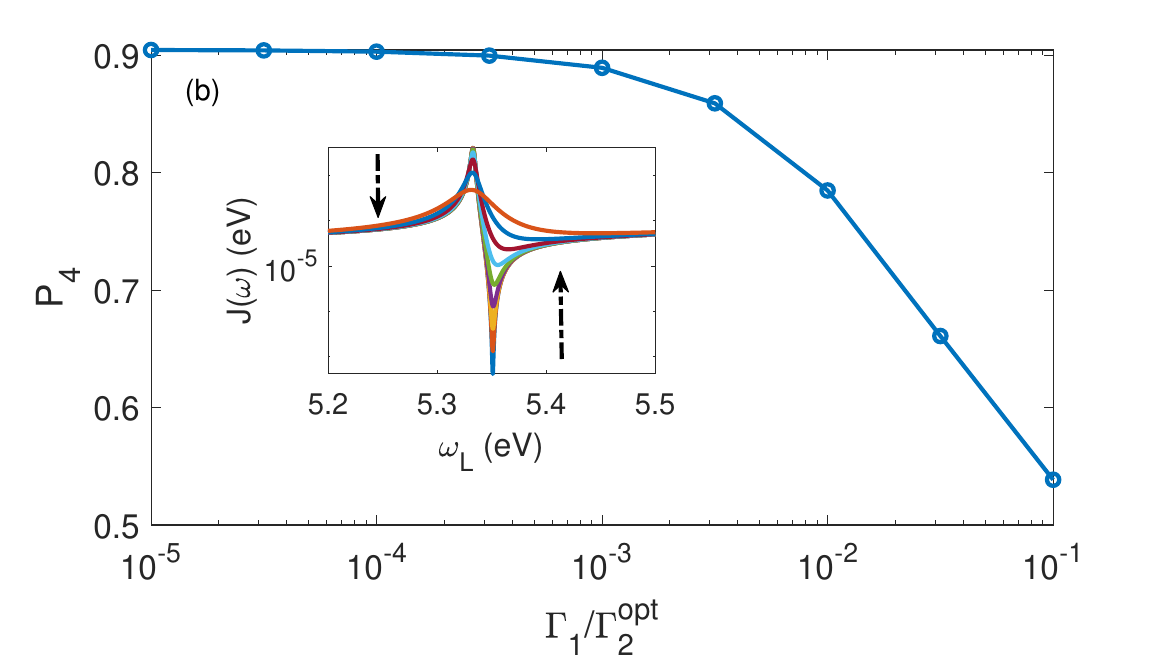}
\caption[]{$P_{4}$ as a function of  $g_1$ (a) and $\Gamma_1$ (b), the coupling strength between the photonic mode and the molecular electronic excitation and the photonic decay rate, respectively. In the insets, the two-mode spectral density is presented for different values of $g_1$ and $\Gamma_1$, where the arrows indicate their trends. When increasing $g_1$, the dip shifts, while for very large $g_1$, the peak disappears. When increasing  $\Gamma_1$,  the peak becomes lower and the dip shallower, and as a result, the spectral density is broader.}
\label{fig:7}
\end{figure}

Finally, we examine the effects of the parameters of the narrow mode, specifically its coupling strength $g_1$ and its decay rate $\Gamma_1$, which are both set to $0$ in Figs.~\ref{fig:3}-\ref{fig:6}. 
We present $P_{4}$ for  $\Omega=10^{-5}$ a.u.  in \autoref{fig:7}(a) as a function of $g_1$ when $\Gamma_1=0$, and in \autoref{fig:7}(b) as a function of $\Gamma_1$ when $g_1=0$. The insets in Figs.~\ref{fig:7}(a)-(b) present the spectral densities for each value of $g_1$ and $\Gamma_1$. 
Increasing $g_1$ leads to a strong coupling with the narrow mode, shifting the dip in the spectral density and thereby affecting the energy selectivity of the cavity. However, this effect is not significant for values of $g_1$ much lower than $g_2^{\mathrm{opt}}$, the coupling strength to the plasmonic mode given in \autoref{tab}. 
Only when $g_1$ reaches $g_2^{opt}/10$, the dip shifts sufficiently for the narrow peak in the spectral density to disappear (see the inset of \autoref{fig:7}(a)), leading to the destruction of cavity-enhanced population transfer to $\Phi_{4}(X)$ (\autoref{fig:7}(a)).
In contrast, the effect of increasing $\Gamma_1$ is much more pronounced (\autoref{fig:7}(b)). Higher $\Gamma_1$ leads to a lower peak and a shallower dip (see inset of \autoref{fig:7}(b)), resulting in a broader peak and reduced energy selectivity. 
Therefore, as shown in \autoref{fig:7}(b), to achieve significant interference between the two modes and to obtain a high $P_4$, $\Gamma_1$ should be $3-4$ orders of magnitude lower than $\Gamma_2^{\mathrm{opt}}$, the decay rate of the plasmonic mode given in \autoref{tab}. 

\section{Summary}
We have explored how hybrid metallodielectric cavities can control photoisomerization reactions at the single-molecule level. These reactions can be tailored via the cavity-induced Purcell effect, which enhances relaxation from the electronically excited state to the desired geometry. In contrast to previous works~\cite{felicetti2020photoprotecting,wellnitz2020collective,torres2021molecular,wellnitz2021quantum} which studied this effect in one-mode plasmonic cavities, we demonstrate that interference between two modes in a hybrid cavity plays a crucial role in achieving energy-selective Purcell enhancement. 
This selectivity leads to higher photoisomerization yields compared to those of one-mode cavities or cavity-free setups.
To illustrate this, we have examined the excited-state asymmetric proton transfer in (Z)-3-aminoacrylaldehyde. Based on electronic-structure calculations and using rate equations that incorporate cavity effects, we study population transfer from the vibronic ground state (proton located on the nitrogen atom) to the fourth excited vibrational state (proton located on the oxygen atom). Optimized, yet realistic, parameters for one-mode plasmonic and two-mode hybrid cavities were used.
Our results reveal that while both cavity setups accelerate the proton transfer reaction and reduce the required laser intensity, the hybrid cavity achieves a significantly higher population transfer. This enhancement comes from the superior energy selectivity of the hybrid cavity, which only targets the desired relaxation pathway. 
We have also analyzed the sensitivity of these results against variations in the parameters of the hybrid cavity and found that is quite robust.
These findings highlight the potential of hybrid cavities for controlling photochemical reactions, paving the way for future applications in cavity-induced chemistry.

\begin{acknowledgments}
This project has received funding from the European Research Council (ERC) under the European Union’s Horizon 2020 research and innovation program (grant agreement no. 852286).
Support from the Swedish Research Council (Grant No.~VR 2022-05005) is acknowledged. In addition, this project received funding from the European Union's Horizon 2020 research and innovation programme through agreement 101070700 (MIRAQLS), as well as under the Marie Skłodowska-Curie Grant Agreement No 101034324. Furthermore, this work was funded by the Spanish Ministry of Science, Innovation and Universities-Agencia Estatal de Investigación through grants PID2021-125894NB-I00, EUR2023-143478 and CEX2023-001316-M (through the María de Maeztu program for Units of Excellence in R\&D).
\end{acknowledgments}

\section*{Author Declaration Section}
\subsection*{Conflict of Interest Statement }
The authors have no conflicts to disclose.

\subsection*{Author Contributions}
\textbf{Anael Ben-Asher}: Conceptualization (equal); Data curation (lead); Investigation (lead);  Visualization (lead); Funding
acquisition (lead); Writing – original draft (lead); Writing – review \& editing (equal). \textbf{Thomas Schnappinger}: Conceptualization (equal); Investigation (equal);  Data curation (supporting);  Writing – original draft (supporting); Writing – review \& editing (equal). \textbf{Markus Kowalewski}: Conceptualization (equal); Funding
acquisition (lead); 
Project administration (lead); Supervision (equal); Writing – original draft (supporting); Writing – review \& editing (equal).
\textbf{Johannes Feist}: Conceptualization (equal); Funding
acquisition (lead); 
Project administration (lead); Supervision (equal); Writing – original draft (supporting); Writing – review \& editing (equal).

\subsection*{Data Availability}
The data that support the findings of this study are available from the corresponding authors upon reasonable request.

\appendix*
\section{Derivation of the rate constant $k_{i\to f}^{ESPT}$ from the Lindblad master equation}

Here, we derive \autoref{kes} from the Lindblad master equation associated with $\hat{H}_0$, treating perturbatively the interaction with the laser and the population transfer to the electronic ground state.
This derivation naturally yields the decay operators for the spontaneous emission of the molecule and the cavity losses, in agreement with Refs.~\citenum{moiseyev2004non,ben2017boomerang,ben2020quantum}. 

The Lindblad master equation associated with the effective NH Hamiltonian $\hat{H}_0$ (\autoref{H}) that also considers the driving by the laser, $\hat{V}_{L}(t)=\Omega D(X)(e^{-i\omega_L t}\sigma_++e^{+i\omega_L t}\sigma_-)$, can be written as 
\begin{equation}
    \frac{d\rho(t)}{dt} = -\frac{i}{\hbar}\big[\hat{H}(t)\rho(t) - \rho(t)\hat{H}^\dagger(t)\big] +\sum_k\hat{V}_d^{(k)}\rho(t)\hat{V}_d^{(k)\dagger},
\end{equation}
where $\hat{H}(t) = \hat{H}_0+\hat{V}_L(t)$ and the sum over $k$ includes the jump operators of the Lindblad superoperator corresponding to the cavity losses and the spontaneous emission of the molecule, such that $\hat{V}_d^{(1)}=\sqrt{\kappa}\sigma_-$, $\hat{V}_d^{(2)}=\sqrt{\Gamma_1}a_1$ and $\hat{V}_d^{(3)}=\sqrt{\Gamma_2}a_2$. 

We consider the eigenstates of $\hat{H}_0$ and $\hat{H}_0^\dagger$, given by $\hat{H}_0|n)=\tilde{E}_n|n)$ and $\hat{H}_0^\dagger|n^*)=\tilde{E}_n^*|n^*)$, respectively, as the non-perturbative zeroth-order states where the notation $|\dots)$ and $(\dots|$ rather than $|\dots\rangle$ and $\langle\dots|$ is used for the right and left eigenstates of NH Hamiltonians, and $\tilde{E}_n$ is complex. Because $\hat{H}_0$ and $\hat{H}_0^\dagger$ are complex symmetric, $|n^*)$ and $(n^*|$ are the complex conjugates of $|n)$ and $(n|$. Furthermore, the states in the zeroth optical excitation manifold of $\hat{H}_0$ and $\hat{H}_0^\dagger$, which can be described by a Hermitian Hamiltonian, obey $|n^*)=|n)=|n\rangle$ and $(n^*|=(n|=\langle n|$ and are real. 
We express the density matrix $\rho(t)$ in terms of these zeroth-order states as $\rho(t)=\sum_{m,n}b_{mn}(t)|m)(n^*|$, and obtain the master equation for the coefficient $b_{mn}=(m|\rho(t)|n^*)$:
\begin{multline}
    \frac{d b_{mn}(t)}{dt}=-\frac{i}{\hbar}\bigg((\tilde{E}_m-\tilde{E}_n^*)b_{mn}(t) \\
    + \sum_l (m|\hat{V}_L(t)|l)b_{ln}(t)+\sum_p(p^*|\hat{V}_L(t)|n^*)b_{mp}(t)\bigg) \\ 
    + \sum_k\sum_{l,p}(m|\hat{V}_d^{(k)}|l)(p^*|\hat{V}_d^{(k)\dagger}|n^*)b_{lp}(t).\label{db}
\end{multline}

Next, we isolate the effect of the laser interaction and Lindblad jump operators by transforming \autoref{db} to the NH interaction picture, as in Ref.~\citenum{ben2024memory}. This yields the following master equation:
\begin{multline}\label{dbI}
    \frac{d \tilde{b}_{mn}(t)}{dt}=-\frac{i}{\hbar}\bigg(\sum_l (m|\hat{V}_L(t)|l)e^{i(\tilde{E}_m-\tilde{E}_l)t/\hbar}\tilde{b}_{ln}(t) \\
    + \sum_p(p^*|\hat{V}_L(t)|n^*)e^{i(\tilde{E}_p^*-\tilde{E}_n^*)t/\hbar}\tilde{b}_{mp}(t)\bigg) + \sum_k\sum_{l,p}\tilde{b}_{lp}(t)\times \\
    (m|\hat{V}_d^{(k)}|l)e^{i(\tilde{E}_m-\tilde{E}_l)t/\hbar}(p^*|\hat{V}_d^{(k)\dagger}|n^*)e^{i(\tilde{E}_p^*-\tilde{E}_n^*)t/\hbar},
\end{multline}
where the coefficients of the density matrix in the interaction picture are given by  $\tilde{b}_{mn}(t)=e^{i(\tilde{E}_m-\tilde{E}_n^*)t/\hbar}b_{mn}(t)$. 
We treat \autoref{dbI} perturbatively by iteratively solving it up to the second order with the initial condition $\tilde{b}_{mn}(0)=\delta_{im}\delta_{in}$. Note that $|i\rangle=|i)=|i^*)$ is the initial state of the population transfer process corresponding to the zeroth optical excitation of $\hat{H}_0$ and $\hat{H}_0^\dagger$. 
As a result, we obtain that the elements of the density matrix in the first optical excitation manifold of $\hat{H}_0$ and $\hat{H}_0^\dagger$ obey:
\begin{multline}\label{bmn}
    \tilde{b}_{mn}(t)=\frac{i}{\hbar}\Omega^2\langle i|D(X)\sigma_-|n^*)(m|D(x)\sigma_+|i\rangle\times \\
    \int_0^tdt'\bigg(\frac{e^{i(\tilde{E}_m-\tilde{E}_n)t'/\hbar}-e^{i(\tilde{E}_m-E_i-\hbar\omega_L)t'/\hbar}}{E_i+\hbar\omega_L-\tilde{E}_n^*} \\
    + \frac{e^{i(\tilde{E}_m-\tilde{E}_n)t'/\hbar}-e^{i(E_i+\hbar\omega_L-\tilde{E}_n^*)t'/\hbar}}{\tilde{E}_m-E_i-\hbar\omega_L}\bigg),
\end{multline}
where $E_i$ is the energy of the initial state and is real. 

Finally, we substitute \autoref{bmn} into \autoref{dbI} and neglect higher-order terms of $\Omega$ to obtain a master equation for the population in the final state $b_{ff}(t)$:
\begin{multline}
    \frac{d {b}_{ff}(t)}{dt}=\Omega^2\sum_k\sum_{l,p}\frac{\langle i|D(X)\sigma_-|p^*)(p^*|\hat{V}_d^{(k)\dagger}|f \rangle}{E_i+\hbar\omega_L-\tilde{E}_p^*} \\
    \times \frac{\langle f|\hat{V}_d^{(k)}|l) (l|D(x)\sigma_+|i\rangle}{\tilde{E}_l-E_i-\hbar\omega_L}   
    \bigg(1+e^{i(\tilde{E}_p^*-\tilde{E}_l)t/\hbar} \\
    - e^{i(E_i+\hbar\omega_L-\tilde{E}_l)t/\hbar}-e^{i(\tilde{E}_p^*-E_i-\hbar\omega_L)t/\hbar}\bigg).\label{dbf}
\end{multline}
Since $|f\rangle=|f)=|f^*)$ is a state within the zeroth optical excitation subspace of $\hat{H}_0$ and $\hat{H}_0^\dagger$, it is associated with the real eignvalue $E_f$ and $b_{ff}(t)=\tilde{b}_{ff}(t)$.
As $t\to\infty$, the time-dependent exponents in \autoref{dbf} vanish due to the positive imaginary part of $\tilde{E}_p^*$ and the negative imaginary part of $\tilde{E}_l$. Since $(l|D(x)\sigma_+|i\rangle=[\langle i|D(X)\sigma_-|k^*)]^*$ and $\langle f|\hat{V}_d^{(k)}|l)=[(l^*|\hat{V}_d^{(k)\dagger}|f \rangle]^*$, this results in the rate constant $k_{i\to f}^{ESPT}=\frac{d {b}_{ff}(t\to\infty)}{dt}$ given in \autoref{kes}.  

%

\end{document}